\documentclass[12pt]{article}
\usepackage{amsfonts}
\usepackage{amsmath}
\usepackage{amssymb}
\usepackage{bm}
\usepackage{dcolumn}
\usepackage{epsfig}
\usepackage{graphicx}
\usepackage{graphics}
\usepackage[latin1]{inputenc}
\usepackage{latexsym}
\usepackage{rotating}
\usepackage{hyperref}

\newcommand\be{\begin{equation}}
\newcommand\ba{\begin{eqnarray}}
\newcommand\ee{\end{equation}}
\newcommand\ea{\end{eqnarray}}

\usepackage{a4wide}
\usepackage{amssymb}
\begin{document}
{\renewcommand{\thefootnote}{\fnsymbol{footnote}}
\hfill  IGC--yy/m--n\\
\medskip
\begin{center}
{\LARGE  Electric Time in Quantum Cosmology }\\
\vspace{1.5em}
Stephon Alexander,$^1$\footnote{e-mail address: {\tt stephon.alexander@dartmouth.edu}}
Martin Bojowald,$^2$\footnote{e-mail address: {\tt bojowald@gravity.psu.edu}}
Antonino Marcian\`o$^1$\footnote{e-mail address: {\tt antonino.marciano@dartmouth.edu}}
and David Simpson$^2$\footnote{e-mail address: {\tt dsimpson@gravity.psu.edu}}\\
\vspace{0.5em}
$^1$ Department of Physics and Astronomy\\
Dartmouth College\\
6127 Wilder Laboratory
Hanover, NH 03755-3528, USA\\
\vspace{0.5em}
$^2$ Institute for Gravitation and the Cosmos,\\
The Pennsylvania State
University,\\
104 Davey Lab, University Park, PA 16802, USA\\
\vspace{1.5em}
\end{center}
}

\setcounter{footnote}{0}

\begin{abstract}
\noindent 
 Effective quantum cosmology is formulated with a realistic global internal
 time given by the electric vector potential. New possibilities for the
 quantum behavior of space-time are found, and the high-density regime is
 shown to be very sensitive to the specific form of state realized. 
\end{abstract}

\section{Introduction}

Even in simple isotropic models, the high-curvature regime of (loop) quantum
cosmology remains poorly understood. At high curvature one expects strong
quantum effects sensitive to what state the universe is in, but the precise
form of suitable states is unknown. (In this paper, we will show a new
explicit example for this sensitivity.) The popular use of Gaussians or
semiclassical states is hard to justify in this regime, and if one starts with
a semiclassical state at low curvature and evolves toward larger curvatures,
the quantum state depends on the dynamics in all cosmic phases passed
through. Quantum ambiguities then prevent precise-enough knowledge of the
state dynamics.

As a second important issue, the problem of time
\cite{KucharTime,Isham:Time,AndersonTime} remains unsolved, affecting the
right choice of dynamics. The problem is usually evaded (but not solved) by
using specific choices of global internal times which tend to be unrealistic
near the big bang, such as a free massless scalar or dust. As part of the
problem of time, it is not known how to transform quantum wave functions or
entire Hilbert spaces between different internal times, and therefore results
found with one choice of global internal time do not necessarily hold for
other choices. But if they depend on what time is used, they cannot be
considered physical.

As elsewhere in physics, effective equations provide a better handle on
reliable predictions. For a canonical setting such as canonical quantum
cosmology, such equations do not refer to entire wave functions but rather to
moments of a state \cite{EffAc,Karpacz}. Only a small number of moments,
chiefly fluctuations and the covariance, is needed in semiclassical regimes,
and as one evolves toward stronger quantum regimes, one can self-consistently
check when higher moments become relevant. Effective techniques for quantum
constraints \cite{EffCons,EffConsRel,EffConsComp} also allow the use of
realistic local internal times, and one can change between different times by
mere gauge transformations \cite{EffTime,EffTimeLong,EffTimeCosmo}. In this
article, we will not make use of these latter techniques because they require
rather involved discussions of constrained systems. Instead, we develop the
effective framework of quantum cosmology for a new choice of internal time
which is still global but more realistic at high density than a free massless
scalar or dust: radiation.  In this article we demonstrate that a more realistic choice of internal time arises from electric fields\footnote{The use of electric fields in early universe cosmology has been implemented in inflationary theories \cite{Ford,Muk,Jab,AMS}}.

\section{Radiation Hamiltonian and effective dynamics}

The Hamiltonian constraint for spatially flat isotropic
Friedmann--Lema\^{\i}tre--Robertson--Walker models with radiation is
\begin{equation} \label{H}
 H=-\frac{3}{8\pi G\gamma^2} c^2 \sqrt{|p|}+ \frac{E^2}{\sqrt{|p|}}=0
\end{equation}
with canonical gravitational variables $(c,p)$ with $|p|=a^2$ and $c=-\gamma
\dot{a}{\rm sgn}(p)$ (with the Barbero--Immirzi parameter $\gamma$ relevant in
loop quantizations \cite{AshVarReell,Immirzi}) in terms of the scale factor
$a$, the derivative being by proper time. We have Poisson brackets
$\{c,p\}=8\pi\gamma G/3$ while the momentum $A$ of $E$, $\{A,E\}=1$, does not
appear in the Hamiltonian. The matter part is determined by the electric field
$E=|\vec{E}|$,\footnote{The absolute value is taken with the flat Euclidean
  metric, $|\vec{E}|=\delta_{ab}E^aE^b$, to keep the spatial metric $q_{ab}$
  as a physical degree of freedom independent of $E$.} assumed sufficiently
small so as not to cause significant anisotropy. By writing the constraint in
the form of a Friedmann equation, dividing $H$ by $|p|^{3/2}$, one can easily
confirm that the matter term provides the correct behavior for radiation: the
$E$-term amounts to an energy density $\rho=E^2/p^2=E^2/a^4$ with $E$ constant
because $H$ does not depend on the momentum $A$ conjugate to
$E$. (Alternatively, the Hamiltonian can be derived using the standard
electric-field energy density: $q_{ab}E^aE^b/\sqrt{\det q}$ reduces to
$|p|E^2/|p|^{3/2}$ with an isotropic spatial metric $q_{ab}=|p|\delta_{ab}$.)

The momentum of $E$ is the electromagnetic vector potential and would
contribute a non-zero term to $H$ in the presence of a magnetic
field. However, a magnetic field requires deviations from homogeneity for the
rotation of $\vec{A}$ to be non-zero. In the symmetric context used here, the
restriction to pure electric fields is therefore meaningful.  Since the
electric field is canonically conjugate to the vector potential $A$, which
does not appear in the constraint, $E$ is constant and $A$ can be used as a
global internal time. We will call this choice {\em electric time}. To realize
$A$-evolution, we follow standard techniques of deparameterization and solve
the constraint equation for the momentum
\begin{equation} \label{pA}
 p_A=E(c,p)=\pm\sqrt{\frac{3}{8\pi G\gamma^2}} |c|\sqrt{|p|}\,.
\end{equation}
As a function on the gravitational phase space $(c,p)$, $E(c,p)$ provides
Hamiltonian equations of motion for the classical $c(A)$ and $p(A)$,
\[
 \frac{{\rm d}c}{{\rm d}A}= \{c,E(c,p)\} \quad \mbox{and} \quad \frac{{\rm
     d}p}{{\rm d}A} = \{p,E(c,p)\}\,,
\]
as well as the basis for the quantum Hamiltonian of effective equations with
respect to $A$. To transform equations or solutions to proper time $\tau$, we
can multiply all ${\rm d}/{\rm d}A$ by ${\rm d}A/{\rm d}\tau= \{A,H\}=
2E/\sqrt{|p|}$, using (\ref{H}). We confirm the correct classical equations of
motion
\[
 \frac{{\rm d}a}{{\rm d}\tau}= \frac{{\rm sgn}(p)}{2\sqrt{|p|}} \frac{{\rm
     d}p}{{\rm 
     d}\tau}= \frac{E}{p} \frac{{\rm d}p}{{\rm d}A}= \frac{E}{p}\{p,E\}=
 \mp \sqrt{\frac{8\pi G}{3}} \frac{E{\rm sgn}(cp)}{\sqrt{|p|}}= -{\rm sgn}(p)
 \frac{c}{\gamma} 
\]
substituting $c$ for $E$ in the last step, and
\begin{eqnarray*}
 \frac{1}{a} \frac{{\rm d}^2a}{{\rm d}\tau^2}&=& - \frac{{\rm sgn}(p)}{\gamma a}
 \frac{{\rm d}c}{{\rm d}\tau}= - \frac{2E}{\gamma p} \frac{{\rm d}c}{{\rm
     d}A}= - \frac{2E}{\gamma p} \{c,E\}= \mp\frac{E}{\gamma p}
 \sqrt{\frac{8\pi G}{3}} \frac{|c|{\rm sgn}(p)}{\sqrt{|p|}}\\
&=& -\frac{8\pi G}{3}
 \frac{E^2}{a^4}= -\frac{4\pi G}{3} (\rho+3P)
\end{eqnarray*}
where we have substituted $E$ for $c$ in the second line and used the
electromagnetic expressions for energy density $\rho$ and pressure
$P=\frac{1}{3}\rho$ to compare with the standard acceleration equation.  (In
what follows, we will set $8\pi G/3=1$ and $\gamma=1$, so that $\{c,p\}=1$.)

\subsection{Effective dynamics}

The sign in (\ref{pA}) determines whether one considers solutions of positive
or negative frequency with respect to time $A$. Without loss of generality, we
will use the negative choice, such that $E(c,p)=-|c|\sqrt{|p|}$. Moreover, we
can choose a definite sign of $p$ (the orientation of space as measured by a
triad) because we will consider only the approach to small $p$, not a possible
transition from positive to negative $p$, or vice versa. For such a transition
in dynamical terms to be described reliably, the Planck regime of quantum
gravity would have to be much better understood than is possible at
present. (To describe the transition non-singularly, we would have to refer to
wave functions subject to a difference equation in loop quantum cosmology
\cite{cosmoIV,IsoCosmo,Sing}.)  We will work with positive $p>0$. Finally,
since $E(c,p)$ is a conserved quantity, the sign of $c\sqrt{p}$ never changes
dynamically and we can drop the absolute value, the two sign options here
merging with the explicit $\pm$ in (\ref{pA}). Even for quantum states, the
fact that the $A$-Hamiltonian $E(c,p)$ and its quantization are conserved
means that the absolute value can be dropped, provided that the expectation
value $\langle\widehat{c\sqrt{p}}\rangle$ is much larger than its quantum
fluctuations. Truncating the whole state to a support of definite sign on the
spectrum not just of $|\widehat{c\sqrt{p}}|$ but also of $\widehat{c\sqrt{p}}$
then ensures that no opposite-sign solutions mix. (These sign issues are the
same as in harmonic cosmology obtained with a free massless scalar
\cite{BouncePert,BounceCohStates}. For more details, see these papers or
\cite{Springer}. For the construction of corresponding Hilbert spaces in
deparameterized quantum cosmology, see \cite{Blyth}.)

Following the procedure of canonical effective equations \cite{EffAc,Karpacz},
the quantum Hamiltonian $E_Q$ is a function on the quantum phase space with
coordinates given by expectation values and moments
\begin{equation}
 \Delta(c^ap^b)= \langle(\hat{c}-\langle\hat{c}\rangle)^a
 (\hat{p}-\langle\hat{p}\rangle)^b\rangle_{\rm symm}
\end{equation}
of a state (using totally symmetric ordering). These variables allow a Poisson
structure by extending
\begin{equation}
 \{\langle\hat{A}\rangle,\langle\hat{B}\rangle\} =
 \frac{\langle[\hat{A},\hat{B}]\rangle}{i\hbar}\,,
\end{equation}
defined for all operators $\hat{A}$ and $\hat{B}$, to products of expectation
values by the Leibniz rule. For fluctuations $(\Delta c)^2=\Delta (c^2)$,
$(\Delta p)^2=\Delta (p^2)$ and the covariance $C_{cp}=\Delta(cp)$, we have
\begin{equation}
 \{(\Delta c)^2,(\Delta p)^2\}= 4C_{cp}\quad,\quad \{(\Delta c)^2,C_{cp}\}=
 2(\Delta c)^2\quad,\quad \{(\Delta p)^2,C_{cp}\}= -2(\Delta p)^2\,.
\end{equation}

Semiclassical states are defined generally by the hierarchy
$\Delta(c^ap^b)\sim O(\hbar^{(a+b)/2})$ of the moments. This class of states
is much more general than that of Gaussian wave functions, which would
determine all moments in terms of at most two parameters. At the level of
effective equations, sufficient generality of the states considered can thus be
guaranteed, without giving rise to prejudices about the form of wave
functions. With a semiclassical (or other) hierarchy, the set of infinitely
many moments can be truncated to finitely many ones by approximation, allowing
practical methods to study the approach to strong quantum regimes. High orders
of the moments, if required, make the equations unwieldy, but the derivation
as well as solutions of equations of motion for moments to rather high orders
can be done with efficient computational codes \cite{HigherMoments}. Quantum
corrections by higher moments are analogs of higher time derivatives in
effective actions \cite{HigherTime}, amounting in quantum cosmology to
important higher-curvature corrections.

The quantum Hamiltonian is a power series in the moments of $c$ and $p$,
obtained by Taylor expanding the quantized $\langle\hat{E}\rangle= \langle
E(\langle\hat{c}\rangle+(\hat{c}-\langle\hat{c}\rangle),
\langle\hat{p}\rangle+(\hat{p}-\langle\hat{p}\rangle))\rangle$ in
$\hat{c}-\langle\hat{c}\rangle$ and $\hat{p}-\langle\hat{p}\rangle$:
\begin{equation}
 E_Q:=\langle\hat{E}\rangle= E(\langle\hat{c}\rangle,\langle\hat{p}\rangle)+
 \sum_{a,b} \frac{1}{a!b!} \frac{\partial^{a+b}
   E(\langle\hat{c}\rangle,\langle\hat{p}\rangle)}{\partial
   \langle\hat{c}\rangle^a \partial \langle\hat{p}\rangle^b} \Delta(c^ap^b)\,.
\end{equation}
Since we assume an operator for $\hat{c}$ to exist, we will obtain the quantum
Hamiltonian of a Wheeler--DeWitt quantization, as opposed to a loop
quantization where only exponentials $\widehat{\exp(i\delta c)}$ exist, but no
$\hat{c}$ \cite{LivRev,Springer}. (A modification of the classical dynamics by
holonomy corrections would be required in the latter case.)  Choosing
totally symmetric ordering for $\widehat{c\sqrt{p}}$ and expanding to
quadratic terms with second-order moments, we have
\begin{equation} \label{EOM}
 E_Q= -c\sqrt{p}- \frac{C_{cp}}{2\sqrt{p}}+ \frac{1}{8} \frac{(\Delta
   p)^2}{p^{3/2}} c +\cdots \,,
\end{equation}
abbreviating $c=\langle\hat{c}\rangle$ and $p=\langle\hat{p}\rangle$ without
risk of confusion. To this order, the Poisson structure provides effective
equations
\begin{eqnarray}
  \frac{{\rm d}p}{{\rm d}A} &=& \sqrt{p}- \frac{1}{8} \frac{(\Delta
    p)^2}{p^{3/2}}\label{p}\\
  \frac{{\rm d}c}{{\rm d}A} &=& -\frac{c}{2\sqrt{p}}+ \frac{C_{cp}}{4p^{3/2}}-
  \frac{3}{16} \frac{(\Delta p)^2}{p^{5/2}}c\\
  \frac{{\rm d}(\Delta p)^2}{{\rm d}A} &=& \frac{(\Delta
    p)^2}{\sqrt{p}} \label{Dp}\\ 
  \frac{{\rm d}C_{cp}}{{\rm d}A} &=& \frac{1}{4} \frac{(\Delta p)^2}{p^{3/2}}c\\
  \frac{{\rm d}(\Delta c)^2}{{\rm d}A} &=& -\frac{(\Delta c)^2}{\sqrt{p}}+
  \frac{1}{2}\frac{cC_{cp}}{p^{3/2}} \label{Dc}
\end{eqnarray}
derived from Hamiltonian equations of motion ${\rm d}f(c,p,\Delta(\cdot))/{\rm
  d}A= \{f,E_Q(c,p,\Delta(\cdot)\}$.

\subsection{Solutions}

The equations (\ref{p}) and (\ref{Dp}) for $p$ and $(\Delta p)^2$ are not
coupled to the other variables and can be solved separately. To do so, we
introduce a new evolution parameter $x$ by ${\rm d}x= p^{-1/2}{\rm d}A$, so
that (\ref{Dp}) is completely decoupled: ${\rm d}(\Delta p)^2/{\rm d}x=(\Delta
p)^2$ is solved by 
\begin{equation} \label{Dpx}
 (\Delta p)^2(x)= (\Delta p)^2_0 e^x
\end{equation}
with initial values
at $x=0$. Inserting this solution in (\ref{p}) and rewriting it for $p^2$, we
have the inhomogeneous differential equation ${\rm d}p^2/{\rm d}x=
2p^2-\frac{1}{4}(\Delta p)_0^2e^x$ solved by
\begin{equation} \label{px}
 p(x)= p_0 e^x \sqrt{1-\frac{1}{4}\frac{(\Delta p)_0^2}{p_0^2}
   (1-e^{-x})}\,.
\end{equation}
The function is real and positive for all semiclassical values, and in fact in
the whole range of $(\Delta p)_0\leq 2p_0$. If $(\Delta p)_0> 2p_0$, in which
case we have to be much more careful trusting our effective equations but may
still analyze (\ref{px}) to suggest possible effects to be corroborated
further, $p(x)$ remains real only for $x < - \ln (1- 4 p_0^2/ (\Delta
p)_0^2)$.

\subsubsection{Potential effects at large fluctuations}

In this brief section we collect properties of our equations and solutions
when fluctuations become large, keeping in mind that we would have to go to
higher orders in effective equations to justify the
implications. Nevertheless, it is interesting to see what the equations may
indicate.

The first derivative of $p$ by $x$,
\[
 \frac{{\rm d}p}{{\rm d}x} = e^{x/2}\, \frac{  \left(p_0^2-\frac{1}{4} (\Delta
     p)_0^2\right)e^x+ \frac{1}{8} (\Delta p)_0^2  
   }{\sqrt{\left(p_0^2-\frac{1}{4} (\Delta p)_0^2\right)e^x+ \frac{1}{4}
       (\Delta p)_0^2}}\,,  
\]
becomes zero at 
\[
 \bar{x}=\ln \left( \frac{\frac{1}{8} (\Delta p)_0^2 }{ -p_0^2 \,+
 \,\frac{1}{4} (\Delta p)_0^2} \right)\,,
\]
which is real and finite for $(\Delta p)_0^2 > 4 p_0^2$. Turning points ---
bounces or recollapses --- therefore require strong quantum effects and large
relative fluctuations.

We can see the option for turning points directly from the equations of
motion, especially (\ref{p}). When fluctuations become large, the moment terms
in the quantum Hamiltonian provide new possibilities for potential turning
points of $p(A)$. We have ${\rm d}p/{\rm d}A=0$ for $(\Delta p)^2=8p^2$,
requiring large relative volume fluctuations. To test whether this point can
be a minimum, potentially corresponding to a bounce, we compute the second
derivative of $p$ by $A$,
\begin{equation} \label{D2p}
 \frac{{\rm d}^2p}{{\rm d}A^2} = \frac{1}{2} - \frac{3}{128} \frac{(\Delta
   p)^4}{p^4} 
\end{equation}
using the equations of motion. (All terms linear in $(\Delta p)^2$ cancel.)
For $(\Delta p)^2=8p^2$, ${\rm d}^2p/{\rm d}A^2=-1/4$, indicating a maximum of
$p$ and therefore a recollapse.  However, if $(\Delta p)^2$ in (\ref{D2p}) is
significant, higher moments may easily contribute and change the behavior. The
precise form of the turning point found here is therefore an explicit example
for an effect that is highly sensitive to the precise form of quantum
state. For other examples in terms of wave functions, see the solutions given
in \cite{GaussianBohmQC,NonSingBohmQC}.

Quantum fluctuations could therefore trigger a recollapse, reminiscent of the
effect pointed out in \cite{EffRecollapse,AltEffRecollapse}. At the turning
point, using our solution (\ref{px}), we have
\[
 p(\bar{x})=
\frac{\Delta p_0}{4\sqrt{1 -4p_0^2/ (\Delta p)_0^2}}>\frac{1}{2}p_0 \,.
\]
The recollapse value $p(\bar{x})$ may be large if $(\Delta p)_0^2$ is
close to $4 p_0^2$, and it is equal to $p_0$ for $(\Delta p)_0^2 =8
p_0^2 $.  (If $(\Delta p)_0^2 =8 p_0^2 $, $x< \ln 2$ in order for
$p(x)$ to be real.)

We can now transform back to $A$ by integrating 
\[
 {\rm d}A=\sqrt{p}{\rm d}x=
 \sqrt{p_0} e^{x/2} \sqrt[4]{1-\frac{1}{4}\frac{(\Delta p)_0^2}{p_0^2}
  (1-e^{-x})}\: {\rm d}x\,.
\]
The result can be expressed in terms of hypergeometric
functions. For instance, using for illustrative purposes the value $(\Delta
p)_0^2=8p_0^2$, we find
\be \label{hypgeo}
 A(x)= A_0- 2\, \sqrt{p_0}\  e^{\frac{x}{4}} \ \sqrt[4]{2 e^x -1} 
\left[ 2- e^x + \sqrt[4]{2} (2- e^x )^{\frac{3}{4}} 
\frac{\Gamma(\frac{5}{4})}{\Gamma(\frac{3}{4}) 
\Gamma(\frac{1}{2})} \ \int_0^x \frac{{\rm d}t}{t^{\frac{1}{4}} 
\ (1-t)^{\frac{1}{2}} \ (1-t\, e^x)^{\frac{1}{4}} }  \right]  \,,
\ee
in which $\Gamma$ represents the Euler function.

Unfortunately, this function and especially its inversion for $x(A)$, to be
inserted in $p(x)$, are complicated. We can proceed further with additional
approximations. First, as one example to explore the strong quantum regime, we
may assume that the initial value $(\Delta p)_0^2$ is close to $8 p_0^2$. Thus
$\bar{x} \rightarrow 0$, and if we are interested in what happens close to
$x\sim 0$, we can expand
\[
 \frac{{\rm d}A}{
{\rm d}x }\sim \sqrt{p_0} \left(1 - \frac{1}{4} x^2
\right) + O(x^3) \,.
\]
Integrating from $0$ to $x$ with
$A_0=0$, we find
\begin{equation}
 \label{aap} A(x)\sim \sqrt{p_0}
x\left(1-\frac{1}{12} x^2
\right) + O(x^4) \,, 
\end{equation}
inverted by
\[
 x= \frac{A}{\sqrt{p_0}}\left(1 + \frac{1}{12} \left( \frac{A}{\sqrt{p_0}}
   \right)^2\right)  + O(A^4)\,.
\]
We can then go back to (\ref{px}), expand it as
$p(x)=p_0(1-\frac{1}{2}x^2)+O(x^3)$,  and find
\[
 p(A)= p_0 \left(1- \frac{1}{2}  \left( \frac{A}{\sqrt{p_0}} \right)^2
 \right)+ O(A^3)\,. 
\]

\subsubsection{Small fluctuations}

For small relative initial fluctuations $(\Delta p)_0^2/p_0^2\ll1$, for which
our effective equations are reliable, and sufficiently small $x$, we can
expand the square root and integrate ${\rm d}A\sim \sqrt{p_0} e^{x/2}
\left(1-\frac{1}{16} ((\Delta p)_o^2/p^2_0)(1-e^{-x})\right){\rm d}x$ to
\begin{eqnarray*}
 A(x)&\sim& 2\sqrt{p_0} \left(\left(1-\frac{1}{16} \frac{(\Delta p)_0^2}{p_0^2}
    \right)e^{x/2}- \frac{1}{16}\frac{(\Delta p)_0^2}{p_0^2} 
e^{-x/2} \right)+A_0\\
&=& 2\sqrt{p_0} \sqrt{1-b^2} \sinh(x/2+{\rm arcosh} (1/\sqrt{1-b^2}))+A_0
\end{eqnarray*}
with $b=1-\frac{1}{8} (\Delta p)_0^2/p_0^2$.  To express $x$ in terms of $A$,
we write
\begin{eqnarray*}
 x&=&2{\rm arsinh}\left( \frac{1}{2} \left(1-(1-\frac{1}{8}(\Delta 
         p)_0^2/p_0^2)^2\right)^{-1/2}(A-A_0)/\sqrt{p_0}
       \right)\\
&&-{\rm arcosh}\left(\left(1-
       (1-\frac{1}{8}(\Delta p)_0^2/p_0^2)^2\right)^{-1/2}\right)\,.
\end{eqnarray*}
We emphasize that we had to assume relative fluctuations to be small only at
one time, which could be in a semiclassical regime. Our solutions are then
valid even in stronger quantum regimes (until, of course, higher moments grow
large).

Finally, we can integrate all equations (\ref{p})--(\ref{Dc}) perturbatively
if we assume moments to be small throughout the whole evolution. At zeroth
order, we first solve the classical equations, ignoring all moments. We obtain
\begin{equation}
 p_{\rm classical}(A)= (\sqrt{p_0}+A/2)^2\quad \mbox{and}\quad c_{\rm
   classical}(A)=\frac{c_0\sqrt{p_0}}{\sqrt{p_0}+A/2}
 \propto \frac{1}{\sqrt{p_{\rm
       classical}(A)}}
\end{equation} 
with initial values $p_0$ and $c_0$ when $A=0$. These solutions can then be
assumed in the equations of motion for moments to find approximate solutions
for the latter. We obtain 
\begin{equation}
 (\Delta p)^2(A)\propto p(A)\quad, \quad C_{cp}(A)\propto
-c(A)+{\rm const}\quad \mbox{and} \quad (\Delta c)^2(A) \propto c^4+{\rm
  const}'c^3+ {\rm   const}'' c^2\,.
\end{equation}
(The first of these equations is, of course, consistent with our full
solutions (\ref{Dpx}) and (\ref{px}) for small $(\Delta p)_0^2/p_0^2$.)  In
particular, relative fluctuations $(\Delta p)^2/p^2\propto p^{-1}$,
$C_{cp}/(cp)\propto p^{-1}$ and $(\Delta c)^2/c^2\propto {\rm const}''$
remain small at small $c$. Moreover, the uncertainty product $(\Delta
c)^2(\Delta p)^2$ is bounded from below by ${\rm const}''$, and the
uncertainty relation will never be violated if we choose appropriate values
for the constants.

These solutions indicate that effective equations get better and better when
one evolves toward larger $p$, but give rise to strong quantum effects with
growing relative fluctuations at small $p$.

\section{Effective equations in terms of proper time}
\label{s:Proper}

The transformation from internal times to proper time after quantization may
not be obvious because it requires a careful look at quantum corrections of
different but related expressions --- the internal-time Hamiltonian and the
Hamiltonian constraint. Nevertheless, once an internal time has been chosen,
there is a unique procedure to transform equations or solutions back to proper
time. We will first go through the general procedure to highlight ambiguities
and difficulties of deparameterized quantizations, as part of the problem of
time.

To obtain equations of motion in proper time in our classical deparameterized
model, we transformed ${\rm d}/{\rm d}A$ to ${\rm d}/{\rm d}\tau=
2E/\sqrt{p}{\rm d}/{\rm d}A$ with constant $E$, computing ${\rm d}A/{\rm
  d}\tau=\{A,H[N]\}$ with a Hamiltonian constraint $H$ of lapse function
$N=1$. In the deparameterized quantum model, we quantize $E(c,p)$ after
having solved the Hamiltonian constraint equation for $E$; we do not quantize
$H$ itself. After deparameterized quantization, we can go back to a
Hamiltonian constraint
\begin{equation}
 H_Q= \frac{E^2}{\sqrt{p}}- \frac{E_Q(c,p,\Delta(\cdot))^2}{\sqrt{p}}=
 \frac{E^2}{\sqrt{p}}- 
 \frac{(c\sqrt{p}+\frac{1}{2}C_{cp}p^{-1/2}- \frac{1}{8}(\Delta p)^2
   cp^{-3/2}+\cdots)^2}{\sqrt{p}}
\end{equation}
with quantum corrections in $E_Q(c,p,\Delta(\cdot))$. By construction, this
corrected Hamiltonian constraint gives rise to the deparameterized model we
started with if the momentum $A$ of $E$ is chosen as time. We therefore
transform $A$-derivatives to proper-time derivatives by using ${\rm d}A/{\rm
  d}\tau=\{A,H_Q\}=2E/\sqrt{p}\approx 2E_Q(c,p,\Delta(\cdot))/\sqrt{p}$, the
latter equation holding on shell when $H_Q=0$.

With this procedure, the relationship between internal-time intervals ${\rm
  d}A$ and proper-time intervals ${\rm d}\tau$ in terms of $E$ does not differ
from the classical one, except that a new function $E_Q$ is used. The
identification of $E$ with $E_Q$, without further quantum corrections, comes
about because $A$ has been chosen as internal time. As a time variable, it is
not quantized and retains its classical form in quantum evolution
equations. By construction, the momentum $p_A$ of $A$ at the quantum level is
then $E_Q$, which is used in the electromagnetic Hamiltonian to compute the
relation between $A$ and proper time.

If we had quantized the Hamiltonian constraint and then solved it or
deparameterized it {\em after} quantization, additional quantum corrections
with moment terms from $E^2/\sqrt{p}$ would have resulted. These corrections
are not included in a deparameterized quantization or in a quantization of the
corresponding reduced phase space. A single deparameterized quantization is
consistent as long as one does not ask how it may be related to other possible
deparameterizations with other choices of internal time. However, if one
allows for different internal times, the resulting quantum theories are not
likely to be equivalent: In each case, one first solves the classical
Hamiltonian constraint equation $H=0$ for a different variable, $E$ in the
choice made here or some other degree of freedom in a different model. The
resulting internal-time Hamiltonians then take into account quantum
corrections which differ from each other and from the corrections that a
quantization of the original Hamiltonian constraint would
imply. Deparameterized quantizations not only ignore some quantum corrections
in the Hamiltonian constraint, the terms ignored even differ depending on
which internal time one chooses. Predictions of different models are then
unlikely to be equivalent, and a single deparameterized model cannot be
considered physical unless one can show how its results are related to
those of different time choices. Our considerations in this paper, as well as
most results in quantum cosmology that rely on quantization after
deparameterization, therefore cannot be considered complete. Our aim here is
not to derive complete effects but rather to give examples for the different
features that various choices of internal time can give rise to. 

After these cautionary remarks, we now continue with our discussion of
electric time. Multiplying our previous equations with $2E_Q/\sqrt{p}$, we
obtain
\begin{eqnarray}
 \frac{{\rm d}p}{{\rm d}\tau} &=& E_Q\left(2- \frac{1}{4} \frac{(\Delta
   p)^2}{p^2}\right)= -2c\sqrt{p}- \frac{C_{cp}}{\sqrt{p}}+ \frac{1}{2}(\Delta
p)^2 \frac{c}{p^{3/2}}\\
\frac{{\rm d}c}{{\rm d}\tau} &=& E_Q\left(-\frac{c}{p}+ \frac{C_{cp}}{2p^2}-
\frac{3}{8} \frac{(\Delta p)^2}{p^3}c\right)=\frac{c^2}{\sqrt{p}}+\frac{1}{4}
(\Delta p)^2 \frac{c^2}{p^{5/2}}\\
\frac{{\rm d}(\Delta p)^2}{{\rm d}\tau} &=& 2E_Q\frac{(\Delta
  p)^2}{p}=-2(\Delta p)^2 \frac{c}{\sqrt{p}}\\
\frac{{\rm d}C_{cp}}{{\rm d}\tau} &=& \frac{1}{2} E_Q \frac{(\Delta
  p)^2}{p^2}c= -\frac{1}{2}(\Delta p)^2 \frac{c^2}{p^{3/2}}\\
\frac{{\rm d}(\Delta c)^2}{{\rm d}\tau} &=& E_Q\left(-2\frac{(\Delta c)^2}{p}+
\frac{cC_{cp}}{p^2}\right)= 2(\Delta c)^2 \frac{c}{\sqrt{p}}- 
C_{cp} \frac{c^2}{p^{3/2}}
\end{eqnarray}
where, consistent with our approximation, we have ignored quadratic terms in
the small second-order moments because they would compete with higher-order
moments which are ignored here. As one can check explicitly, the moments
satisfy
\begin{equation}
\frac{\rm d}{{\rm d}\tau}\left((\Delta p)^2(\Delta c)^2- C_{cp}^2\right)=0
\end{equation}
so that the uncertainty product is preserved. Any departure of a state
from saturating the uncertainty relation
\begin{equation} \label{Uncertainty}
 (\Delta p)^2(\Delta c)^2- C_{cp}^2\geq \frac{\hbar^2}{4}
\end{equation}
remains constant. If an initial state saturates the uncertainty relation, it
will keep saturating it in this regime, and we obtain a dynamical coherent
state.

From the first of these equations it follows that the classical relation
$c=-\frac{1}{2}\dot{p}/\sqrt{p}= -\dot{a}$, which descends from the isotropic
reduction of the Ashtekar--Barbero connection $A_a^i=\Gamma_a^i-\gamma K_a^i$
($\gamma=1$) for a spatially flat cosmological model, receives quantum
corrections: writing $p$ in terms of the scale factor, we have
\begin{equation}
c= -\frac{\dot{a}+\frac{1}{2}C_{cp}/p}{1-\frac{1}{4}(\Delta p)^2/p^2}=
-\dot{a}\left(1+\frac{1}{4}(\Delta p)^2/p^2\right)-
 \frac{1}{2}C_{cp}/a^2+\cdots\,.
\end{equation}
(We cannot easily transform the moments to those of $a$ and $\dot{a}$ because
these variables are non-linearly related to $p$ and $c$.)
With this relation, we can use the equation for ${\rm d}c/{\rm d}\tau$ to
compute the acceleration equation
\begin{equation} \label{acc}
 \frac{\ddot{a}}{a}= -\left(1+\frac{1}{4}\frac{(\Delta p)^2}{p^2}\right)
 \left(\frac{\dot{a}}{a}\right)^2\,,
\end{equation}
ignoring products of moments. For small relative fluctuations, the equation is
consistent with the classical one for radiation. Only the $p$-fluctuation
enters, while all covariance terms cancel. Even if fluctuations could be
large, they would not cause acceleration.

It is interesting to analyze these equations near the time when
$\dot{a}=0=\dot{p}$ (which, as we recall, requires large fluctuations), where
we have small $c=-\frac{1}{2}C_{cp}/a^2$ and can ignore $c^2$-terms in our
equations. In addition to $a$, $c$ is then nearly constant and we can
integrate the moment equations
\begin{equation}
\frac{{\rm d}(\Delta p )^2}{{\rm d}\tau} = -2(\Delta p)^2 \frac{c}{a}
 \quad,\quad
\frac{{\rm d}C_{cp}}{{\rm d}\tau} = 0 \quad,\quad
\frac{{\rm d}(\Delta c)^2}{{\rm d}\tau} = 2(\Delta c)^2\frac{c}{a}
\end{equation}
to obtain
\begin{equation} \label{cpExp}
 (\Delta p)^2(\tau)\sim (\Delta p)_1^2 \exp(-2(\tau-\tau_1)c/a) \quad,\quad
 (\Delta c)^2(\tau)\sim (\Delta c)_1^2 \exp(2(\tau-\tau_1)c/a)
\end{equation}
where $\tau_1$ is the time when $\dot{a}=0$.

\section{Numerical solutions}

We return to the Hamiltonian system in (\ref{EOM}), to seek
self-consistent solutions. Here we only illustrate differences between
classical and quantum solutions and test the validity of our
approximations, which turn out to be very good. A detailed analysis
especially near potential turning points could, as motivated by
\cite{AK}, reveal a wave function of the Universe that entails parity
violation and non-trivial chiral effects. However, not just the
numerics but also higher orders of our effective equations would then
have to be developed.  

We attempt to solve the system (\ref{EOM}) with certain initial
conditions for $c(A)|_{A=A_0}$ and the other variables.  We may assume
the normalization of $p(A)$ initially, such that $p(A)|_{A=A_0}=1$. We
then fix $C_{cp}|_{A=A_0}=C_{cp}^0$, $(\Delta p)^2_{A=A_0}=(\Delta
p)^2_0$, and $(\Delta c)^2_{A=A_0}:=(\Delta c)^2_0$, respecting the
uncertainty relation (\ref{Uncertainty}). Our effective equations are
valid for some time provided that $(\Delta p)^2_0\ll p_0^2$, and so
on, but we will choose larger values just to illustrate the
implications of quantum corrections. Even for $(\Delta p)^2_0\approx
p_0^2$, it turns out that we remain close to the classical solutions
for rather long times.

We therefore solve the system (\ref{EOM}) parametrizing solutions in
terms of two positive real numbers, namely $(\Delta p)^2_0\,, (\Delta
c)^2_0\in \mathbb{R}^+$, and a real number $C_{cp}^0\in \mathbb{R}$,
which together with $p(A)|_{A=A_0}$ and $c(A)|_{A=A_0}$ represent our
initial conditions for the system. The moments must obey the
uncertainty relation (\ref{Uncertainty}). With these initial values,
we solve (\ref{p})--(\ref{Dc}).

\begin{figure}[tb]
\begin{center}
 \vspace{-2mm}
\!\includegraphics[scale=1]{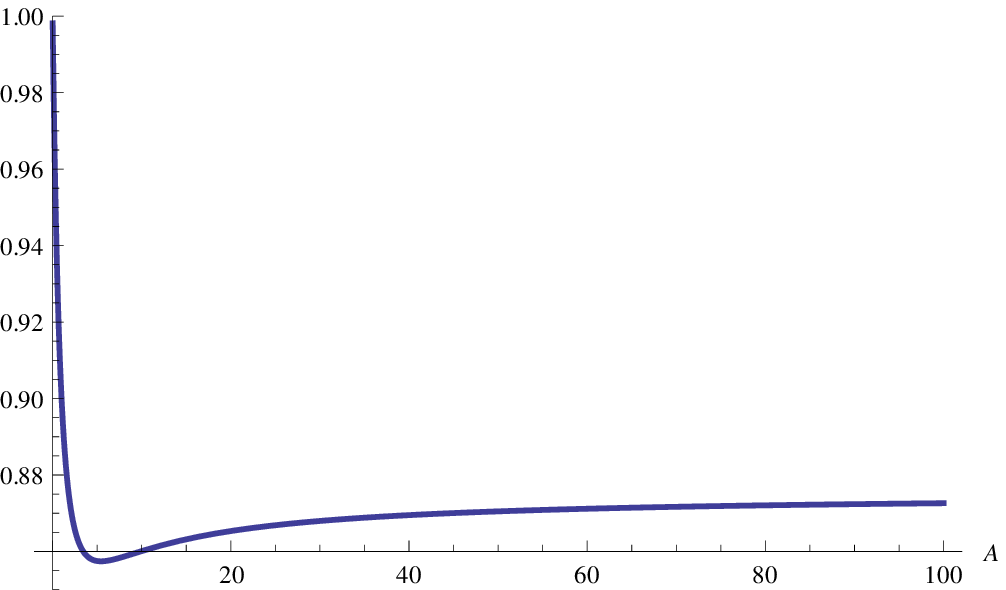} \hspace{4mm}
\includegraphics[scale=1]{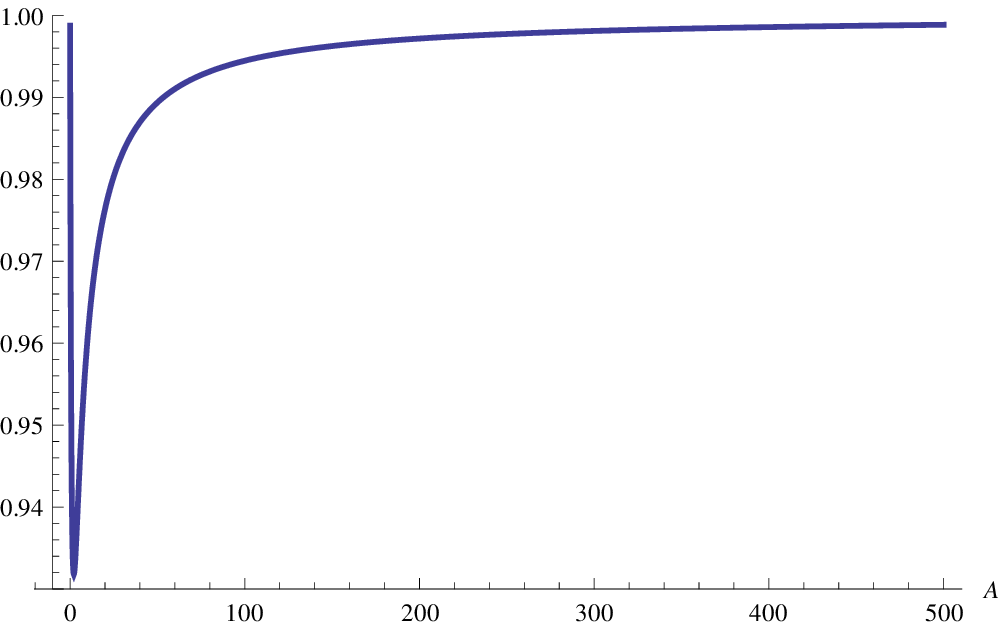}
\vspace{-2mm}
\caption{Solutions of effective equations for expectation values, plotted as
  their ratios to the classical solutions $c_{\rm classical}(A)$ (top) and
  $p_{\rm classical}(A)$ (bottom). Initial fluctuations have been set to
  rather large values --- $\Delta p_0=\Delta c_0=p_0=c_0=1$ with $C_{cp}=0$
  --- to show the implications of quantum corrections more
  clearly. Nevertheless, the ratios stay close to one. \label{f:cpratios}}
\end{center}
\vspace{-2mm}
\end{figure}

\begin{figure}[tb]
\begin{center}
 \vspace{-2mm}
\!\includegraphics[scale=1]{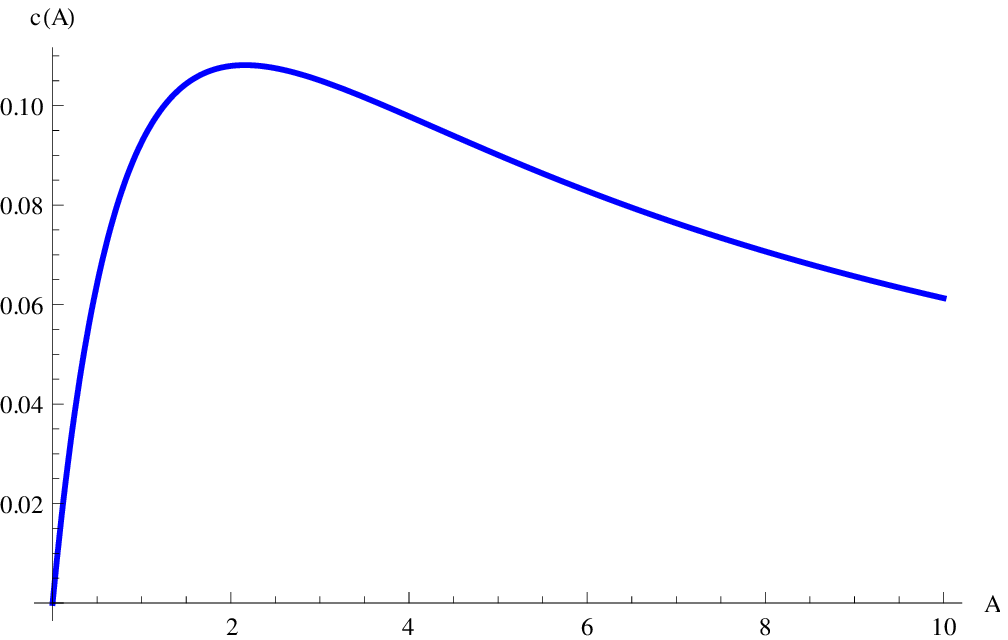} \hspace{4mm}
\includegraphics[scale=1]{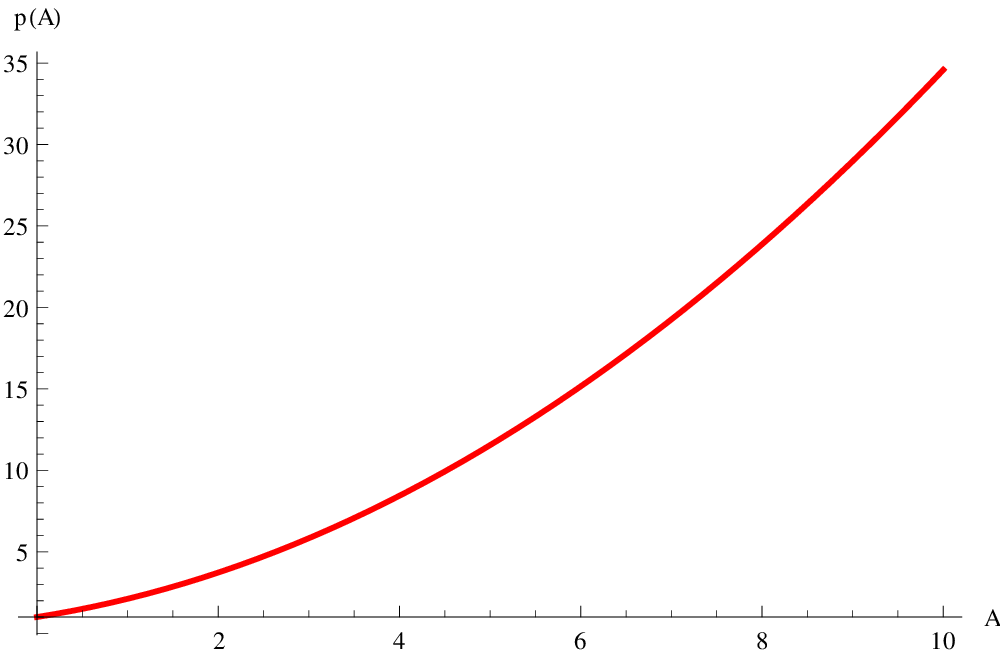}
\vspace{-2mm}
\caption{The dynamics in $A$ of the expectation values on the quantum state of
  the universe of $c(A)$ and $p(A)$, computed by solving the second-order
  equations (\ref{p})--(\ref{Dc}) numerically. Here, $c_0=0$ initially, and
  $C_{cp}=3/4$, all other initial values as in
  Fig.~\ref{f:cpratios}. \label{f:cp}}
\end{center}
\vspace{-2mm}
\end{figure}

\begin{figure}[tb]
\begin{center}
\vspace{-2mm}
\includegraphics[scale=1.19]{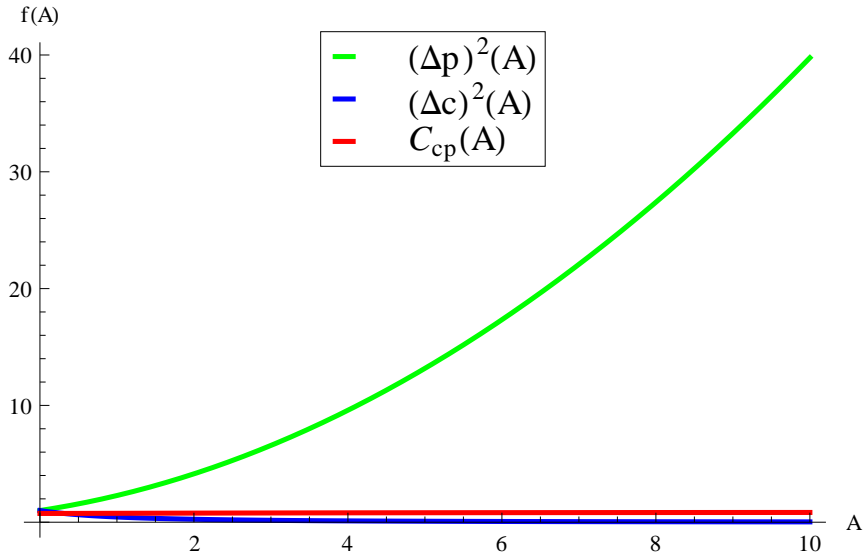}
\vspace{2mm}
\caption{The variance of $p$ (in green), of $c$ (in blue), and the
covariance between the two operators (in red) for a quantum state of
the universe that minimizes the uncertainty relation
(\ref{Uncertainty}). Good agreement for instance with (\ref{cpExp}) is
realized. (Initial values as in Fig.~\ref{f:cp}.) \label{f:xyz}}
\end{center}
\vspace{-4mm}
\end{figure}

\begin{figure}[tb]
\begin{center}
\vspace{-2mm}
\includegraphics[scale=1.19]{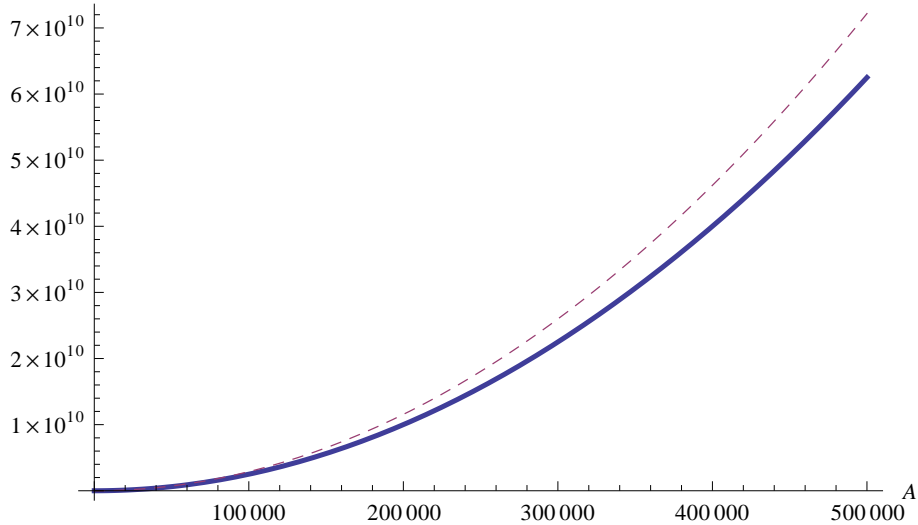}
\vspace{2mm}
\caption{The squared $p$-variance $(\Delta p)^2(A)$ (dashed)
compared with the expectation value $p(A)$ (solid). Both functions
increase in nearly the same way even for large initial fluctuations as
in Fig.~\ref{f:cpratios}, comfirming our analytical solutions. We have
$(\Delta p)^2>p$, but relative fluctuations $(\Delta p)^2/p^2$ become very
small at large $A$. 
 \label{f:Dp}}
\end{center}
\vspace{-4mm}
\end{figure}

\begin{figure}[tb]
\begin{center}
\vspace{-2mm}
\includegraphics[scale=1.19]{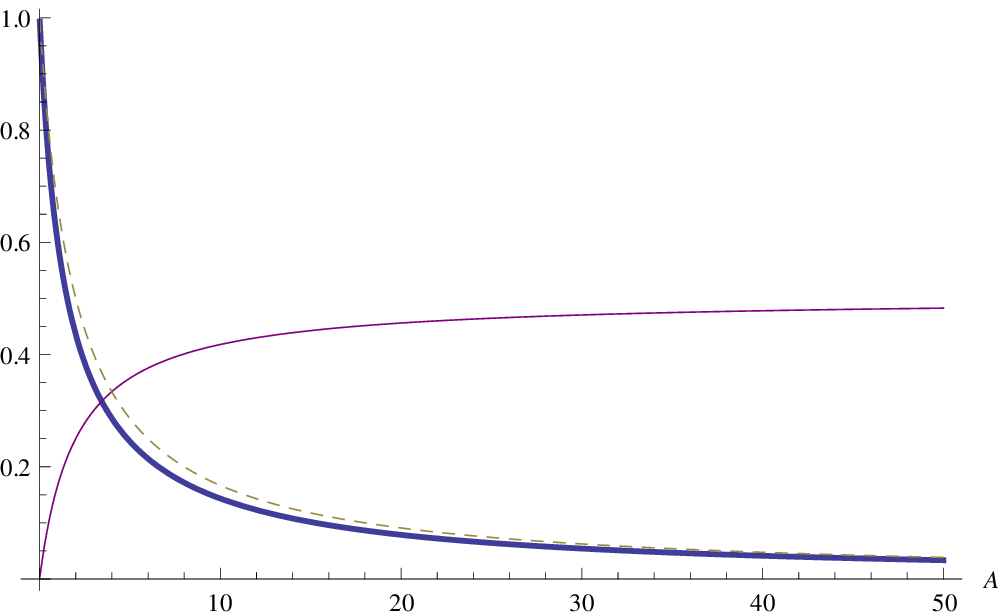}
\vspace{2mm}
\caption{The same qualitative agreement as in Fig.~\ref{f:Dp}, seen
for the expectation value $c(A)$ (thick) and the covariance
$C_{cp}(A)$ (thin). (Also shown is the classical $c_{\rm
classical}(A)$ for comparison.) These graphs agree
with the analytical solution $C_{cp}(A)\propto {\rm const}-c(A)$. 
\label{f:Ccp}}
\end{center}
\vspace{-4mm}
\end{figure}

Figure~\ref{f:cpratios} shows good agreement with the classical solutions even
for rather large fluctuations, with $\Delta p_0=p_0$ and $\Delta c_0=c_0$
($C_{cp}=0$).  Numerical solutions assuming $c_0=0$ together with $(\Delta
p)^2_0 = (\Delta c)^2_0 = 1$ and $C_{cp}^0=3/4$ are shown in
Fig.~\ref{f:cp}. The plot for $p(A)$ is increasing in $A$, with a certain rate
amounting to the expansion of the universe. The plot for $c(A)$ seems to
indicate acceleration in the increasing branch, but this is realized only in
terms of internal time $A$, not in terms of proper time (see (\ref{acc})).

In Fig.~\ref{f:xyz}, we see numerical evolutions of the second-order
moments.  The $c$-variance decreases in $A$ and reaches quite soon a
vanishing value. On the other hand, the $p$-variance increases with
the same rate of $p(A)$. The covariance for the position and momentum
operators is almost stable in the range studied, while the two
fluctuations evolve according to (\ref{cpExp}). The good agreement
with our analytical solutions and approximations is further
illustrated in Figs.~\ref{f:Dp} and \ref{f:Ccp}.

\section{Conclusions}

We have laid out the basic description of deparameterized quantum cosmology
with time provided by the electric field. The only matter source required to
formulate time evolution is radiation, expected to be significant in any
early-universe model. No artificial matter sources such as dust or free
massless scalars are required.

Our discussion in this paper does not include modifications suggested by loop
quantum cosmology, such as holonomy and inverse-triad corrections. The former
would prevent us from using $c$ as an operator and in moments, making the
Hamiltonian more non-linear by replacing $c$ with functions such as
$\sin(\delta c)/\delta$. Such modifications certainly alter the high-curvature
behavior and are expected to compete with fluctuation terms. The magnitude of
holonomy corrections depends on the parameter $\delta$ used in this
modification (a quantization ambiguity). However, if $\delta$ is related to
the Planck length, for instance by $\delta\sim\ell_{\rm P}/\sqrt{p}$ as often
assumed, holonomy corrections are of the size $\ell_{\rm P}^2/\ell_{\cal H}^2$
with the Hubble scale $\ell_{\cal H}^2$, the same size expected for
higher-curvature corrections. Quantum back-reaction by moments, on the other
hand, contributes higher-derivative terms, also related to higher-curvature
corrections. If a loop quantization is used, it is therefore impossible to
study either holonomy corrections or quantum back-reaction by fluctuations in
isolation. In this paper, to provide a manageable first discussion of quantum
cosmology with electric time, we have therefore decided to ignore loop effects
so that our solutions refer to quantum back-reaction in Wheeler--DeWitt
models. (The alternative, ignoring quantum back-reaction but keeping holonomy
corrections, is not consistent unless one restricts oneself to models in which
quantum back-reaction is absent or weak. These are only the models of harmonic
cosmology \cite{BouncePert,Harmonic} or kinetic-dominated ones
\cite{BounceSqueezed,QuantumBounce,HighDens}.) Moreover, recent results in
off-shell loop quantum gravity
\cite{ConstraintAlgebra,JR,ScalarHol,ThreeDeform,TwoPlusOneDef,TwoPlusOneDef2,AnoFreeWeak}
have shown that holonomy corrections in loop quantum cosmology must be treated
with care because they trigger signature change at high density
\cite{Action}. Their evolution equations must therefore stop before
interesting high-density effects can be realized.

Our discussion of the quantum dynamics was done mainly at the effective level
of Wheeler--DeWitt quantum cosmology, allowing us to consider large classes of
states without tying us down to specific wave functions (such as
Gaussians). Considering quantum corrections by fluctuations, we have found
several new possibilities for early-universe dynamics, showing also the high
sensitivity of potential turning points to specific forms of states.  The
small-volume behavior in electric time is rather different from that with a
free massless scalar as time. In the latter case, the singularity is
approached exponentially with respect to the scalar (or by a power law in
proper time) while electric time, which does not give rise to a harmonic
model, leads to a more complicated behavior. Although we have pointed out the
possibility of inequivalent quantum corrections obtained in different
deparameterizations (Section~\ref{s:Proper}), it remains to be seen whether
these differences are solely due to the different matter sources or due to
effects of choices of internal times. To decide this question,
effective-constraint methods \cite{EffCons,EffConsRel} would be suitable but
are rather complicated to perform. We therefore end this paper by concluding
that much remains to be done before the high-curvature regime of quantum
cosmology can be controlled, including a better understanding of the role of
internal times.

\section*{Acknowledgements}

This work was supported in part by NSF grant PHY0748336 and a grant from
the Foundational Questions Institute (FQXi).


\end{document}